# The US Algorithmic Accountability Act of 2022 vs. The EU Artificial Intelligence Act: What can they learn from each other?

Jakob Mökander,[1] Prathm Juneja,[1] David Watson,[2] Luciano Floridi[1,3]

[1] Oxford Internet Institute, University of Oxford, Oxford, OX1 3JS, UK

[2] Department of Statistical Science, University College London, London WC1E 7HB, UK

[3] Department of Legal Studies, University of Bologna, Bologna, 40126, Italy

**Abstract**

On the whole, the U.S. Algorithmic Accountability Act of 2022 (US AAA) is a pragmatic approach to balancing the benefits and risks of automated decision systems. Yet there is still room for improvement. This commentary highlights how the US AAA can both inform and learn from the European Artificial Intelligence Act (EU AIA).

**A tale of two acts**

On February 3rd, Senator Ron Wyden, Senator Cory Brooker and Representative Yvette Clark introduced the Algorithmic Accountability Act of 2022 (US AAA) in the US Senate and the House of Representatives (Office of U.S. Senator Ron Wyden, 2022a). The bill addresses growing public concerns about the widespread use of automated decision systems (ADS). It proposes that organisations deploying such systems must take several concrete steps to identify and mitigate the social, ethical, and legal risks. As a legislative effort to regulate ADS across industries, the US AAA is the latest milestone in a worldwide trend to complement or replace self-regulation in this domain with legislation (Floridi, 2021). The most influential example of that trend is the Artificial Intelligence Act (EU AIA), proposed by the European Commission (2021).

However, the similarities between the US AAA and the EU AIA are only apparent. Consider politics first. The EU AIA was proposed by the EU's executive branch, highlighting strong institutional backing for the act (even if it will probably evolve before passing into legislation). In contrast, the US AAA has yet to win support in the Senate or the House. While the bill is a revised (and improved) version of the 2019 Algorithmic Accountability Act, it remains unclear whether it will gather sufficient political support to become law.  The two acts also differ in style and depth. The EU AIA is a lengthy, sometimes opaque document that attempts to lay down rules for using ADS and provide details about how these are to be enforced. In comparison, the US AAA takes a relatively high-level approach. It defines critical terminology and stipulates requirements that owners of ADS must fulfil. However, it delegates questions concerning implementation to the Federal Trade Commission (FTC).

In short, the US AAA and the EU AIA spring from widely different political contexts and legislative traditions. Despite these differences, comparing both the framing and the content of the two documents offers valuable insights. [1] In previous works, we have highlighted both promising and challenging aspects of the US AAA (see Mökander & Floridi, 2022), and the EU AIA  (see Mökander et al., 2021), respectively. In this commentary, we go one step further by comparing the relative strengths and limitations of the two proposals. We aim to address a simple yet pertinent question: what can the US AAA and the EU AIA learn from each other?

**Promising signs**

The US AAA and the EU AIA have much in common. For example, neither act seeks to prohibit or limit the use of ADS. Instead, they both aim to establish the governance infrastructure needed to hold bad actors accountable and allow actors with good intent to ensure and demonstrate that their ADS are ethical, legal and safe. To that end, the US AAA requires organisations to perform *impact assessments* of (i) ADS before their deployment and (ii) augmented decision-making processes after the deployment of ADS (US AAA, Sect. 3). This two-pronged approach mirrors the *conformity assessments* and *post-market monitoring* plans mandated by the EU AIA.

The inclusion of both ex-ante and ex-post assessments is welcome, insofar as it accounts for the fact that ADS evolve and update their internal decision-making logic over time. However, like all governance mechanisms, impact assessments have limitations. For example, they may fail to identify and mitigate specific harms, or reduce ethics to a box-ticking exercise. Yet, combined with strong institutional backing, impact assessments contribute to creating traceable documentation and help spark ethical deliberation in organisations that design and deploy ADS (Selbst, 2021). Hence, the focus on procedural regularity and transparency in both documents is promising.

The proposed American and European legislations also differ in several ways – three of which are worth mentioning here. The US AAA's primary merit is that it is framed in terms of ADS rather than the more popularised term 'AI systems', preferred by the European Commission. The two terms are often used interchangeably in the literature. However, the term *automated decision systems* better captures the technical features we care about, which rely on a heterogeneous mix of machine learning algorithms and hard-coded argumentation frameworks. By focusing on regulating 'critical decision processes' rather than 'high-risk AI systems', the US AAA avoids the ontological question of what an AI system is *and* accounts for the fact that the level of automation, in decision-making processes, is best understood as a difference of degree on a spectrum. Because the definition of ADS is technology-agnostic, it is also future-proof. Further, framing the legislation in terms of ADS also prevents distracting discussions about the nature of intelligence, cognition or consciousness that are often associated with the term *artificial intelligence.* In short, the US AAA's terminology is both scientifically sound and coherent with its regulatory objective. A revised EU AIA should adopt the same.

A second advantage of the US AAA is the demarcation of its scope. Its transparency obligations apply to companies that 'employ ADS to make critical decisions', i.e., any decision that has significant legal or material effects on a consumer's life. This includes access to education, employment and financial services (US AAA, Sect. 2.7). In contrast, the EU AIA only requires so-called 'high-risk AI systems' to undergo conformity assessment. At first glance, this difference may seem of little importance. However, the shift is significant, since ethical tensions do not emerge from using ADS alone but can also be related to the broader context of ADS-supported decision-making tasks (Danks & London, 2017). Therefore, it makes sense to – as the US AAA does – avoid questions about what an 'AI system' is and focus instead on identifying those decision-making processes that require additional layers of public oversight.

Finally, the US AAA helpfully introduces a benchmark for ethical and legal evaluation by requiring organisations to compare the performance of a new ADS with that of the pre-existing decision-making processes that it is intended to augment or replace. That is reasonable because both human decision-makers and ADS come with their own (complementary) sets of strengths and weaknesses. The risks associated with ADS are well-known and include privacy violations and discriminatory outcomes (Tsamados et al., 2021). At the same time, human judgement is subject to many cognitive biases and can be influenced by prejudices and circumstantial factors (Kahneman et al., 2021). Therefore, when properly used, ADS can lead to more objective and potentially fairer decisions. The US AAA accounts for this dynamic by requiring organisations deploying new ADS to 'describe the existing decision-making process' and 'explain the intended benefits of augmenting [it]' (US AAA, Sect. 4). This requirement helps technology providers and courts compare ADS to the relative affordances and limitations of human decision-makers and subject them to appropriate and proportional quality assurance and transparency obligations.

**Unresolved tensions**

Despite its many merits, the US AAA still leaves room for improvement. Next, we highlight three areas where it can be strengthened by learning from the EU AIA.

First, the US AAA applies only to 'large companies' that either (a) have an annual turnover over $50 million, (b) have over $250 million in equity value, or (c) process the information of over 1 million users. That exclusive focus on companies is unfortunate because many critical (and often automated) decisions are made by government agencies that are outside the FTC's jurisdiction, such as ADS-based tools used by local governments to help determine which families should be investigated by child welfare agencies (Stapleton et al., 2022). Moreover, while the exception for small- and medium-sized enterprises (SMEs) is understandable, its current formulation in the bill is unhelpful. The cost of complying with new regulations tends to impact SMEs disproportionately. So, policymakers should avoid raising the barriers to entry into already highly concentrated markets. However, not subjecting all ADS to the same transparency requirements exposes data subjects to unnecessary risks and could enable malicious actors to avoid regulatory oversight by outsourcing tasks via 'creative accounting'. Although the US AAA mentions that smaller corporations that are 'substantially owned, operated, or controlled' by a 'large company' will have to follow these rules, it is unclear if simply contracting out automated decisions to smaller companies would be covered under the legislation. The EU AIA offers a better model, placing consistent requirements on all ADS but offering targeted support to SMEs to reduce their costs for ensuring and demonstrating compliance (EU AIA, Article 55).

Second, the US AAA seeks to ensure both the equal treatment of decision subjects and equal outcomes for different protected groups, a hardly achievable aspiration. When presenting the bill, Senator Hirono (a co-sponsor of the bill) said that the US AAA 'will require companies to look at the impact of their automation' and that 'consumers deserve fair and equitable treatment.'(Office of U.S. Senator Ron Wyden, 2022b) These are laudable ends, yet values often conflict and require trade-offs. For example, ADS may improve a decision-making process's overall accuracy but risk discriminating against specific subgroups in the population (Whittlestone et al., 2019). Similarly, different definitions of fairness – like individual fairness, demographic parity and equality of opportunity – are mutually exclusive in all but the most trivial problems (Kleinberg, 2018). According to Sect. 4.4 in the US AAA, covered entities are required to 'perform ongoing evaluation of any differential performance associated with data subjects' race, color, sex, gender, age, disability, religion, family-, socioeconomic-, or veteran status... for which the covered entity has information' The issue here stems from the lack of clarity around the word 'performance.' Companies may choose to define performance narrowly, allowing them to obfuscate discriminatory impacts by choosing not to include specific metrics in their definition (e.g., an ADS that gives out the same proportion of loans to people of differing backgrounds, but discriminates in the loan size). Similarly, companies may choose to stop collecting information about protected classes to avoid having to conduct this analysis. Finally, research has shown that adversarial methods can be used to pass standard fairness audits even with algorithms designed to ensure discriminatory impact (Kearns et al., 2018; Slack et al., 2020), suggesting that technical hurdles remain for monitoring and enforcement. It is unclear what weight this requirement carries regarding other performance metrics like accuracy, efficiency and privacy.

Third, the US AAA – like many U.S. bills – is much less specific than the EU AIA. On the upside, the expansion of the bill, from 15 to around 50 pages (from 2019 to 2022), enables adequate definitions of terms such as ADS and 'covered entity' (i.e., a person or organisation to which the bill applies). Still, the US AAA delegates many important choices about policy design to the FTC. For example, the bill stipulates that covered entities must conduct impact assessments, but it is left open for the FTC to determine what documentation and information must be submitted after completing such an assessment. Because its specifics are yet unknown, it is hard to compare the US AAA thoroughly with the EU AIA, which outlines a Europe-wide assurance ecosystem in detail. This is not necessarily a negative; most laws passed by Congress delegate some authority to executive agencies (Clouser McCann & Shipan, 2022). However, the US AAA's lack of specificity is an issue in the instances where it is unnecessarily vague without delegating those details to the FTC. For example, covered entities are required 'to the extent possible [to] consult with stakeholders such as technology experts and representatives of impacted groups' (US AAA, Sect. 3). The (frequent) inclusion of qualifiers such as 'to the extent possible' risks watering down the proposed legislation.

## Towards digital governance

Overall, the US AAA represents a pragmatic approach to the problem of managing the legal and ethical challenges posed by ADS. The framing is sound, the proposed enforcement mechanisms are well-established, and the bill explicitly seeks to balance the required regulatory oversight with incentives for innovation. Further, we acknowledge that many differences between the US AAA and the EU AIA highlighted in this commentary result from differing political processes and legal traditions (Vokinger & Gasser, 2021).

That said, the current American effort to regulate ADS is too modest. After all, not only 'consumers' but 'citizens' in general are increasingly affected by ADS. Moreover, a policy is only as strong as the institutions backing it. While the EU AIA is part of a long-term, holistic effort by the EU to shape the digital ecosystem in the Union and beyond, the US AAA constitutes only a fragmented attempt. Following the introduction of the GDPR in 2016, a so-called 'Brussels effect' has been observed, whereby multinational organisations choose to harmonise all their international data management practices with EU laws for practical reasons (Bradford, 2020). The EU AIA may have a similar effect. In contrast, the impact of the US AAA – if it passes into law – would almost certainly be limited without a more coordinated approach. Given the US's economic weight and technological leadership, it is regrettable that there may not be a 'Washington effect' positively shaping global digital ecosystems.

On the one hand, recent developments suggest that the US is increasing efforts in tech diplomacy. For example, the US and the EU have agreed to develop a joint roadmap on evaluation and measurement tools for trustworthy AI and risk management (*EU-US Trade and Technology Council*, 2022). On the other hand, it is still unclear whether the US AAA itself has any chance of passing, given that recent attempts at US versions of EU technology law, such as the American Innovation and Choice Online Act of 2022 (an analogue to the EU's Digital Markets Act) and the Data Protection Act of 2021 (an analogue to GDPR) have stalled in Congress. The current political landscape in the US is not friendly to new technology regulation, with only two pieces of federal technology legislation passing in the last 25 years (Kang, 2022). Recent polling has shown that public support for more regulation of major technology companies has fallen from 56% of Americans in April 2021 to 44% in May 2022 (Vogels, 2022). This, coupled with a divergence in partisan goals (e.g., the Republicans focus on claims of discrimination against conservatives on social media and the Democrats focus on claims of misinformation (Kang & McCabe, 2022)), makes it unlikely that the US

AAA advances further in Congress. What remains is the possibility of executive action by the Democrats' majority in the FTC, whose focus appears to be on anti-trust and privacy regulation, not the regulation of ADS (Harding McGill & Gold, 2022).

Fundamentally, the US AAA, like the EU AIA, is about more than just ADS. It is about what decisions we should consider critical and what outcomes we should strive towards. The hard questions that need to be addressed concern what decision criteria and evidence (input data) are to be considered legitimate (or at least socially acceptable) for different private and public decision-making processes. Answering such questions requires a positive vision of what future societies should be. Policymakers should therefore move beyond attempts to secure minimal 'algorithmic accountability' and focus instead on designing public governance mechanisms that allow organisations to strike justifiable trade-offs within the limits of legal permissibility and commercial viability to shape how ADS are designed and what ends they serve.

## References

Bradford, A. (2020). The Brussels Effect [Bookitem]. In *The Brussels Effect*. Oxford University Press. https://doi.org/10.1093/oso/9780190088583.003.0003

Clouser McCann, P. J., & Shipan, C. R. (2022). How many major US laws delegate to federal agencies? (almost) all of them. *Political Science Research and Methods*, *10*(2), 438–444. https://doi.org/10.1017/PSRM.2021.32

Danks, D., & London, A. J. (2017). Algorithmic bias in autonomous systems. *IJCAI International Joint Conference on Artificial Intelligence*, *0*(January), 4691–4697. https://doi.org/10.24963/ijcai.2017/654

European Commission. (2021). *Proposal for regulation of the European parliament and of the council - Laying down harmonised rules on artificial intelligence (artificial intelligence act) and amending certain Union legislative acts.*

*EU-US Trade and Technology Council.* (2022, May 16). https://ec.europa.eu/commission/presscorner/detail/en/IP_22_3034

Floridi, L. (2021). The End of an Era: from Self-Regulation to Hard Law for the Digital Industry. *Philosophy and Technology*, *34*(4), 619–622. https://doi.org/10.1007/s13347-021-00493-0

Harding McGill, M., & Gold, A. (2022). *Lina Khan's to-do list on Big Tech*. AXIOS. https://www.axios.com/2022/05/12/ftc-majority-lina-khan-to-do-list-big-tech

Kahneman, D., Sibony, O., & Sunstein, C. R. (2021). *Noise : a flaw in human judgment.*

Kang, C. (2022). *As Europe Approves New Tech Laws, the U.S. Falls Further Behind*. The New York Times. https://www.nytimes.com/2022/04/22/technology/tech-regulation-europe-us.html

Kang, C., & McCabe, D. (2022). *Efforts to Rein In Big Tech May Be Running Out of Time*. The New York Times. https://www.nytimes.com/2022/01/20/technology/big-tech-senate-bill.html

Kearns, M., Neel, S., Roth, A., & Wu, Z. S. (2018). Preventing fairness gerrymandering: Auditing and learning for subgroup fairness. *35th International Conference on Machine Learning, ICML 2018*, *6*, 4008–4016.

Kleinberg, J. (2018). *Inherent Trade-Offs in Algorithmic Fairness*. https://doi.org/10.1145/3219617.3219634

Mökander, J., Axente, M., Casolari, F., & Floridi, L. (2021). Conformity Assessments and Post-market Monitoring : A Guide to the Role of Auditing in the Proposed European AI Regulation. *Minds and Machines*, 1–27.

Mökander, J., & Floridi, L. (2022). From algorithmic accountability to digital governance. *Nature Machine Intelligence 2022*, 1–2. https://doi.org/10.1038/s42256-022-00504-5

Office of U.S. Senator Ron Wyden. (2022a). Algorithmic Accountability Act of 2022. *117th Congress 2D Session*. https://doi.org/10.1016/S0140-6736(02)37657-8

Office of U.S. Senator Ron Wyden. (2022b). *Wyden, Booker and Clarke Introduce Algorithmic Accountability Act of 2022 To Require New Transparency And Accountability For Automated Decision Systems*. https://www.wyden.senate.gov/news/press-releases/wyden-booker-and-clarke-introduce-algorithmic-accountability-act-of-2022-to-require-new-transparency-and-accountability-for-automated-decision-systems

Selbst, A. D. (2021). An Institutional View of Algorithmic Impact Assessments. *Harvard Journal Of Law & Technology*, *35*. https://papers.ssrn.com/sol3/papers.cfm?abstract_id=3867634

Slack, D., Hilgard, S., Jia, E., Singh, S., & Lakkaraju, H. (2020). *Fooling LIME and SHAP: Adversarial Attacks on Post hoc Explanation Methods*. https://doi.org/10.1145/3375627.3375830

Stapleton, L., Cheng, H.-F., Kawakami, A., Sivaraman, V., Cheng, Y., Qing, D., Perer, A., Holstein, K., Steven Wu, Z., & Zhu, H. (2022). *Extended Analysis of "How Child Welfare Workers Reduce Racial Disparities in Algorithmic Decisions."* https://github.com/logan-stapleton/AFST_racial_disparity.

*Text - S.2134 - 117th Congress (2021-2022): Data Protection Act of 2021.* (2021). http://www.congress.gov/

*Text - S.2992 - 117th Congress (2021-2022): American Innovation and Choice Online Act.* (2022). http://www.congress.gov/

Tsamados, A., Aggarwal, N., Cowls, J., Morley, J., Roberts, H., Taddeo, M., & Floridi, L. (2021). The ethics of algorithms: key problems and solutions. *AI & SOCIETY 2021 37:1, 37*(1), 215–230. https://doi.org/10.1007/S00146-021-01154-8

Vogels, E. A. (2022). *More tech regulation seen less favorably in U.S. than it was in 2021*. Pew Research Center. https://www.pewresearch.org/fact-tank/2022/05/13/support-for-more-regulation-of-tech-companies-has-declined-in-u-s-especially-among-republicans/

Vokinger, K. N., & Gasser, U. (2021). Regulating AI in medicine in the United States and Europe. In *Nature Machine Intelligence* (Vol. 3, Issue 9, pp. 738–739). Nature Research. https://doi.org/10.1038/s42256-021-00386-z

Whittlestone, J., Alexandrova, A., Nyrup, R., & Cave, S. (2019). The role and limits of principles in AI ethics: Towards a focus on tensions. *AIES 2019 - Proceedings of the 2019 AAAI/ACM Conference on AI, Ethics, and Society*, 195–200. https://doi.org/10.1145/3306618.3314289